# Movie Recommender Systems: Implementation and Performace Evaluation

**Mojdeh Saadati, Syed Shihab, Mohammed Shaiqur Rahman**
Iowa State University
shihab@iastate.edu
msaadati@iastate.edu
shaiqur@iastate.edu

**Abstract**

Over the years, an explosive growth in the number of items in the catalogue of e-commerce businesses, such as Amazon, Netflix, Pandora, etc., have warranted the development of recommender systems to guide consumers towards their desired products based on their preferences and tastes. Some of the popular approaches for buiding recommender systems, for mining user derived input datasets, are: content-based systems, collaborative filtering, latent-factor systems using Singular Value Decomposition (SVD), and Restricted Boltzmann Machines (RBM). In this project, user-user collaborative filtering, item-item collaborative filtering, content based recommendation, SVD, and neural networks were chosen for implementation in Python to predict the user ratings of unwatched movies for each user, and their performances were evaluated and compared.

## 1  Introduction

Before the industrial age, information age, and globalization, when the number of choices for the consumer in any market was limited, it was possible to solely rely on word of mouth, editorial and staff reviews of movies, books, etc., testimonials, endorsements and general surveys to make the right purchases . However, this natural social approach to gathering recommendations became impractical as the assortment of offerings in the market grew from a few dozens to millions, often overwhelming customers trying to decide what to buy, watch, read, etc. This paved the way for the deployment of intelligent recommender systems by businesses, applying statistical and knowledge discovery techniques on users' purchases datasets, to generate sound product suggestions for the buyers, as a value-added service.

One such company, where a recommender system has now become indispensable, is Netflix. In 2006, Netflix announced a competition for developing recommender systems that can outperform its own system, Cinematch, leading to the succesful application of different techniques such as RBMs for movie rating predictions. To facilitate this, they released a dataset containing 100 million anonymous movie ratings and reported their Root Mean Squared Error (RMSE) performance on a test dataset as 0.9514 (James Bennett, 2007). A standard metric for evaluating the performance of rating predictors and, in general, classifiers (Herlocker, 2004), the RMSE metric is computed using the recommender's prediction, $o_i$, and the actual rating provided by a user, $t_i$, for $n$ such ratings in the test set, as shown in equation 1. If the RMSE is squared, we get the Mean Squared Error (MSE).

$$RMSE = \left[\frac{1}{n}\sum_{i=1}^{n}(o_i - t_i)^2\right]^{1/2} \qquad (1)$$

This paper gives an overview of the key ideas of the recommendation techniques implemented in the project in section 2, followed by a description of their implementation and results in section 3, and finally concludes with a summary in section 4. In this paper, the terms "items" and "movies"are used interchangeably.

### 2.1  Collaborative Filtering

Tapestry, the first recommender system to be produced, used Collaborative Filtering (CF) to aggregate the evaluations (Goldberg, 1992). In traditional CF, each user is represented as a vector of items of size $N_i$, where $N_i$ is the number of unique items being offered by the business (Linden, 2003). The vector entries may be explicit ratings provided by the user for certain items using a certain rating scale, or boolean number 1 for any item purchased by the user and 0 otherwise. Since $N_i$ is very large, this vector is mostly sparse. When all these user rating vectors are put together in a matrix, with rows and columns representing individual users and items respectively, we get the utility or rating matrix of size $N_u \times N_i$. An example of a utility matrix for four users and four movies is given in the table below, where the '?' denote missing entries. The user rating vectors for users 1 and 2, $\vec{r_1}$ and $\vec{r_2}$, are given in equations 2 and 3 respectively.

|        | Movie1 | Movie2 | Movie3 | Movie4 |
|--------|--------|--------|--------|--------|
| User 1 | 5      | -      | 4      | 4      |
| User 2 | -      | 5      | 4      | 1      |
| User 3 | 3      | 3      | -      | -      |
| User 4 | 1      | 4      | -      | 2      |

**Table 1: Rating matrix for four users and four movies**

$$\vec{r_1} = [5 \quad ? \quad 4 \quad 4] \quad (2)$$

$$\vec{r_2} = [? \quad 5 \quad 4 \quad 1] \quad (3)$$

Based on these user vectors, it is possible to compare their likings and dislikings of products with other users. For any given user X, a neighborhood N, comprising of K most similar users, is formed first, such that the users in the neighborhood have similar tastes and preferences to that of user X. An illustration of this process is given in the figure below. The left side of figure has users and right side has products. A solid line from user to product indicates the user has liked that product. A dashed line between users indicates the users are similar and a dashed line from a product to a user indicates that the product has been recommended to that user.

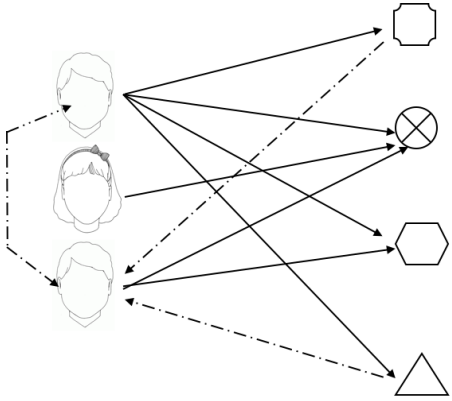

**Figure 1: An illustration of user-user CF**

Various similarity measures have been used to this end, such as Jaccard similarity, cosine similarity, centered cosine similarity, etc. This project used the centered cosine similarity, which is also known as Pearson Correlation. For using the popular centered cosine metric, the ratings are at first normalized by subtracting the row means from each row entries, and then the missing ratings are treated as zero, which is the new average of each row; this helps to capture our intuition of similar users. In other words, the ratings of each user is now "centered" around zero. After following these steps, a modified rating matrix is obtained, such as that given below for the example problem discussed earlier. Next, to calculate the cosine similarity between any two vectors, such as that shown in Figure 1, equation 4 is used.

|        | Movie1 | Movie2 | Movie3 | Movie4 |
|--------|--------|--------|--------|--------|
| User 1 | 0.67   | **0**  | -0.33  | -0.33  |
| User 2 | **0**  | 1.67   | 0.67   | -2.33  |
| User 3 | 0      | 0      | **0**  | **0**  |
| User 4 | -1.33  | 1.67   | **0**  | -0.33  |

**Table 2: Modified rating matrix after normalization of each row and replacement of missing ratings with zeroes**

$$\cos\theta = \frac{\vec{r_1}\cdot\vec{r_2}}{\|\vec{r_1}\|\|\vec{r_2}\|} \quad (4)$$

Two approaches can then be used to estimate the ratings of user X of items he/she has not tried out yet: user-user CF and item-item CF. In user-user CF, the ratings of users in N are used to predict the rating of an item I by user X using either equation 5 to calculate the average rating of the ratings of item I by K neighbors in N; or equation 6 to calculate the weighted average rating of the ratings of item I by neighbors in N, where the weights are the similarity values. Finally, a list of 10 top rated items are recommended to user X.

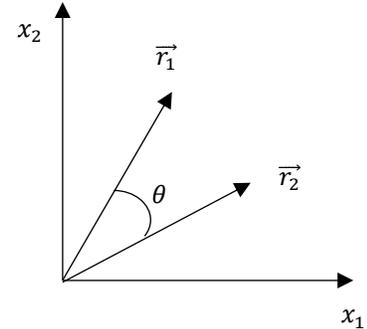

**Figure 2: Two ratings vectors in a 2-D space separated by angle $\theta$**

$$r_{x,M} = \frac{\sum_{y \in N} r_{yM}}{K} \quad (5)$$

$$r_{x,M} = \frac{\sum_{y \in N} s_{xy} r_{yM}}{\sum_{y \in N} s_{xy}} \quad (6)$$

Item-item CF, a dual approach to user-user CF, estimates the rating of an item I by forming a neighborhood of similar items instead of similar users using centered cosine similarity. As before, the unknown rating is estimated by taking the weighted average of the known ratings of the similar items.

## 2.3 Content-based method

While the CF techniques uses only users' ratings data to predict ratings, the content-based recommendation technique uses movie features, such as actors, directors, genre, etc., in addition to ratings data. Each movie is represented as a binary feature vector, such as that shown in table 1, where only two features – actor A and actor B – are considered. In this example, actor A starred in movie 1 and movie 2, and actor B acted in the rest of the movies.

|         | M1 | M2 | M3 | M4 | M5 |
|---------|----|----|----|----|----|
| Actor A | 1  | 1  | 0  | 0  | 0  |
| Actor B | 0  | 0  | 1  | 1  | 1  |

**Table 3: Representation of movies as binary feature vectors**

For all users, a user profile vector is created by normalizing their ratings, and then using the normalized average feature ratings as the vector components. Then, user profile vectors and rated movie feature vectors are compared for each user using centered cosine similarity, followed by using the weighted averaging technique to estimate the unknown ratings.

## 2.2 Neural networks

Neural Network is a collection of connected units or nodes called neurons and the connection between neurons are called synapses which trasmit a signal from one neuron to another.

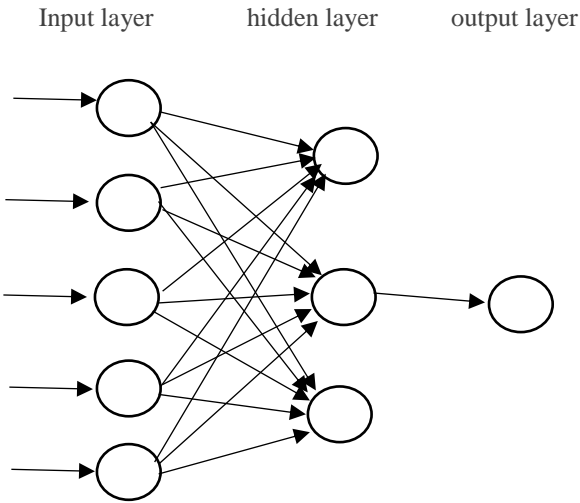

**Figure 3: A sample neural network**

Each neuron is characterized by an a function f(x), which is defined as a composition of other functions $g_i(x)$ that can further be decomposed into other functions. This can be conveniently represented as a network structure, shown in Figure 3, which has 3 layers and 9 units, where the hidden layer outputs depend on the input layer values, input-to-hidden weights, and hidden neuron activation functions.

Equation 7 shows show how neuron outputs are calculated, where K commonly refers to an activation function or "nonlinearity" of a unit. K is some predefined function, such as the hyperbolic tangent, sigmoid function, softmax function, rectified linear function, etc. Neural Networks can be used for both classification and regression problems.

$$f(x) = K(\sum_i w_i g_i(x)) \quad (7)$$

## 2.3 Singular Value Decomposition

SVD is essentially a matrix factorization approach, where the utility matrix is assumed to be formed from a scalar product of two other matrices – the user matrix, $P$ of size $N_u \times k$, and feature matrix, $Q$ of size $N_i \times k$ – as shown in equation 8, where $k$ is a hyperparameter denoting the number of latent factors.

As all the users and items are mapped to a latent $k$-dimesional space, where the axes are the so-called factors, SVD is, in essence, carrying out a dimensionality reduction since $k$ is very small compared to $N_i$ and $N_u$. Once $Q$ and $P$ have been learned, the rating for any item I and user X can be computed by equation 9, which takes the dot product of $q_i$ and $p_x$, the feature vectors of item I and user X respectively.

$$U \cong Q \cdot P^T \quad (8)$$

$$\hat{r}_{xi} = q_i \cdot p_x^T \quad (9)$$

The learning problem of determining matrices Q and P can be reduced to an optimization problem, , as shown in equation 10, where the Sum-of-Squared Error (SSE) is minimized over $Q$ and $P$ to obtain the minimum reconstruction error.

$$\min_{P,Q} \sum_{(i,x) \in R} (r_{xi} - q_i \cdot p_x^T)^2 \quad (10)$$

## 3 Implementation and Results

For this project, the dataset, comprising of 1,000,209 anonymous ratings of approximately 3,900 movies made by 6,040 users, was collected from the MovieLens (Konstan, 2015). Uniques UserIDs and MovieIDs are used to represent individual users and movies respectively. All the users have at least rated 20 movies, on a scale of 1 to 5. For training and testing the recommender systems, the holdout method was used to split the dataset into two parts: training data comprising 75% of the data, and testing data having the remaining 25%.

In processing the data driven from movieLen and imdb database we used the same imdbID variable that exist in movielen data to map these two data set. this mapping helps us to gather both users features from movieLen and movie feature form imdb. After mapping the two dataset, by getting help from imdbpy library that exist in python programming language, we extract all movie features which are, imdb ranking, release year, genres, country that the movie is made in, main actor or actors and director. In the next step to make this main database suitable for neural network algorithm that we used, we did decoding of string varibles for attribues 'genres' and 'country' to numerical ones.

### 3.1 Collaborative Filtering

We used MovieLens dataset and python language to implement this approach. We used ratings.csv file, downloaded from the GroupLens website, to extract ratings of 7120 users for 14026 movies. For user-user CF, we used centered cosine similarities between two users to find which two users match each other in their movie taste. When two users are similar, the similarity value of their ratings is close to 1 (or positive when compared to other users); hence, these two users are likely to watch similar movies. So, if one has not watched a movie the other one has watched and rated high, the system recommends that movie to the first user who has not watched it.

A matrix with rows consisting of users and columns consisting of movies was created with ratings as values. For the cosine calculaiton, we normalized the ratings, making '0' rating as the base reference. This made the average of all the ratings for a user in a row equal to zero. Then we used the cosine values of different user-user combinations to form a neighborhood for each user to calculate the ratings of the movies that the user has not either watched or rated.

The RMSE value obtained using user-user CF was 3.59163265578297. Based on the predicted ratings, the top 10 rated movies that were recommeded to a particular user with userID-3 are:

1. Pulp Fiction (1994)  Comedy | Crime | Drama | Thriller
2. Forrest Gump (1994) Comedy | Drama | Romance | War
3. Braveheart (1995) Action | Drama | War
4. Schindler's List (1993) Drama | War
5. Apollo 13 (1995) Adventure | Drama | IMAX
6. American Beauty (1999) Comedy | Drama
7. Seven (a.k.a. Se7en) (1995) Mystery | Thriller
8. Fargo (1996)  Comedy | Crime | Drama | Thriller
9. Fight Club (1999)  Action | Crime | Drama | Thriller
10. Dances with Wolves (1990) Adventure | Drama | Western

Similarly, we implemented the item-item CF, but here instead of finding two similar users, we tried to find two similar movies (items) and based on the centered cosines values of different item pairs, we predicted the ratings of movies not watched by a user and then recommend top 10 rated movies to that user.

Again, we evaluated the performance using RMSE, which was found to be 3.6361576424237274, in this case.

The top 10 movies recommended to same user (userID-3) were found to be different now as listed below:

1. Michael (2011) Drama|Thriller
2. Your Sister's Sister (2011) Comedy|Drama
3. Tomboy (2011) Drama
4. One I Love, The (2014) Comedy|Drama|Romance
5. Over the Edge (1979) Crime|Drama
6. Kill List (2011) Horror|Mystery|Thriller
7. Undefeated (2011) Documentary
8. Out of the Furnace (Dust to Dust) Drama|Thriller
9. Distant (Uzak) (2002)  Drama
10. Page Turner, The (Tourneuse de pages, La) (2006)  Drama|Musical|Thriller

### 3.2 Content based method

Here, we predicted the ratings based on the features of the movies. We construct a binary matrix wth rows consisting of movies and columns consisting of features based on whether a feature is available in that movie or not.

The features we extracted for the movies were: ActorID-1, ActorID-2, ActorID-3, ActorID-4, ActorID-5, ActorID-6, ActorID-7, ActorID-8, ActorID-9, ActorID-10, ActorID-11, ActorID-12, ActorID-13, ActorID-14, DirID-1, DirID-2, DirID-3, DirID-4, DirID-5, DirID-6, DirID-7, DirID-8, DirID-9, DirID-10, DirID-11, DirID-12, DirID-13, DirID-14, DirID-15, DirID-16, DirID-17, DirID-18, DirID-19, Action, Adventure, Animation, Belgium, Comedy, Crime, Drama, Family, Fantasy, Horror, Music, Mystery, Romance, Sci-Fi, Spain, Thriller, UK, USA, France, Germany. The real names of the actors and directors, abbreviated as "dir", can be found using the IMDB database.

We calculated the cosine similarities between pairs of movies. The ratings of the movies can be predicted using the weighted averaging technique for recommendation purpose.

### 3.3 SVD

We implemented this approach using python language. A matrix with rows as users and columns as movies was created and filled with ratings as values. We imported the desired python packages to run the algorithm.

We evaluated our result based on RMSE values for different values of k, where k is the number of features.

|  | K=3 | K=25 | K=75 | K=99 |
|---|---|---|---|---|
| RMSE | 3.557 | 3.555 | 3.542 | 3.587 |

**Table 4: Results obtained by varying the number of features**

Based on the predicted rating, we recommended top 10 movies to the same user as listed below:

1. Fargo (1996)
2. Schindler's List (1993)
3. Pulp Fiction (1994)
4. Casablanca (1942)
5. Dr. Strangelove or: How I Learned to Stop Worr...
6. Monty Python and the Holy Grail (1975)
7. Groundhog Day (1993)
8. L.A. Confidential (1997)
9. Forrest Gump (1994)
10. American Beauty (1999)

### 3.4 Neural Network

One of the work that we have done in our movie project was to predict the rating that each user will give to specific movies by doing neural network classification of different rating classes. We tried two approaches while using neural networks for predicting movie ratings. In the first approach, we made one neural network for all users. Due to computational issues, we considered only 9 users while building this neural network. In the second approach, we tried implementing one neural network for each individual user. In order to implement neural networks in Python, we used a function MLPClassifier from neural sklearn.neural_network library which gives the parameters, hidden units, hidden layer and activation function as an input to build the neural network. In the next step we feed the network with the training dataset that is obtained from deviding the main dataset into two set of training and testing dataset with user's data of 1 to 9, and in the other time we feed the neural network with the data of user one only. For analysing the acuracy of algorithm we used MSE to compare actual result with the obtained result. The results, presented in the tables below, show the neural network had more accuracy with activation function 'tanh', and in general it works better when one neural network is used for each user.

| Activation | Hidden Layers: 4 Hidden Nodes: 12 | Hidden Layers: 8 Hidden Nodes: 12 | Hidden Layers: 4 Hidden Nodes: 6 |
|---|---|---|---|
|  | MSE | MSE | MSE |
| 'relu' | 1.45 | 1.58 | 0.81 |
| 'logistic' | 0.65 | 0.65 | 0.64 |
| 'identity' | 0.725 | 0.82 | 1.78 |
| 'tanh' | 0.73 | 0.85 | 0.6 |

**Table 5: Results obtained using the approach of employing one neural network for all users**

| Activation | Hidden Layers: 4 Hidden Nodes: 12 | Hidden Layers: 8 Hidden Nodes: 12 | Hidden Layers: 4 Hidden Nodes: 6 |
|---|---|---|---|
|  | MSE | MSE | MSE |
| 'relu' | 2.87 | 0.17 | 1.00 |
| 'logistic' | 0.30 | 0.22 | 0.17 |
| 'identity' | 1.02 | 1.05 | 1.2 |
| 'tanh' | 0.27 | 0.17 | 0.22 |

**Table 6: Results obtained using the approach of employing one neural network made for each user**

### 4 Conclusion

Recommender systems add value to businesses by assisting their customers in selecting the right product out of uncountable choices. This project implemented five of the popular movie recommendation approaches to predict unknown ratings and recommend to users accordingly. After implementation, their performances were compared using RMSE and MSE metrics. SVD was found to have performed better than CF. The variation of $k$, the number of features, in SVD did not change RMSE significantly. Two different approaches were tried out for neural networks: one neural network for all users, and one neural network for each user. In both cases, 'tanh' activation function performed better than others.

### Acknowledgments

We are highly thankful to the Grouplens for making the datasets available to be downloaded over internet which we used for our experiments and we are also very thankful to IMDB for providing API's to fetch data from its database. At the last but not the least, we are thankful to our Professor and all TA's for helping us to understand the concepts.